\documentclass[11pt,a4paper,article,amssymb,amsmath,bm,nofootinbib]{revtex4-1}
\usepackage{graphicx}

\newcommand{\be}{\begin{equation}}
\newcommand{\ee}{\end{equation}}
\newcommand{\bea}{\begin{eqnarray}}
\newcommand{\eea}{\end{eqnarray}}

\begin{document}

\centerline{\huge $K \to \pi$ matrix elements of the}
\vspace{0.25cm}
\centerline{\huge chromomagnetic operator on the lattice}

\vspace{0.75cm}

\centerline{\Large M.~Constantinou$^{(a)}$, M.~Costa$^{(b)}$, R.~Frezzotti$^{(c)}$, V.~Lubicz$^{(d,e)}$,}

\vspace{0.1cm}

\centerline{\Large G.~Martinelli$^{(f)}$, D.~Meloni,$^{(d,e)}$ H.~Panagopoulos$^{(b)}$, S.~Simula$^{(e)}$}

\vspace{0.5cm}

\centerline{\includegraphics[draft=false]{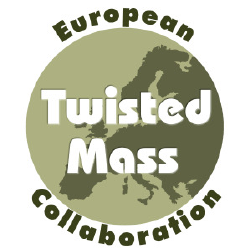}}

\vspace{0.5cm}

\centerline{\it $^{(a)}$Department of Physics, Temple University, Philadelphia, PA 19122-1801, USA}

\centerline{\it $^{(b)}$Department of Physics, University of Cyprus, Nicosia, CY-1678, Cyprus}

\centerline{\it $^{(c)}$Dip. di Fisica, Universit\`a di Roma ``Tor Vergata'' and INFN, Sezione di ``Tor Vergata'',}
\centerline{\it Via della Ricerca Scientifica 1, I-00133 Rome, Italy}

\centerline{\it $^{(d)}$Dipartimento di Matematica e Fisica, Universit\`a Roma Tre, I-00146 Rome, Italy}

\centerline{\it $^{(e)}$INFN, Sezione di Roma Tre, Via della Vasca Navale 84, I-00146 Rome, Italy}

\centerline{\it $^{(f)}$Dip. di Fisica, Universit\`a  di Roma ``La Sapienza" and INFN, Sezione di Roma,}
\centerline{\it P.le Aldo Moro 2, I-00185 Rome, Italy} 

\begin{abstract}
We present the results of the first lattice QCD calculation of the $K \to \pi$ matrix elements of the chromomagnetic operator $O_{CM} = g\, \bar s\, \sigma_{\mu\nu} G_{\mu\nu} d$, which appears in the effective Hamiltonian describing $\Delta S = 1$ transitions in and beyond the Standard Model. Having dimension 5, the chromomagnetic operator is characterized by a rich pattern of mixing with operators of equal and lower dimensionality. The multiplicative renormalization factor as well as the mixing coefficients with the operators of equal dimension have been computed at one loop in perturbation theory. The power divergent coefficients controlling the mixing with operators of lower dimension have been determined non-perturbatively, by imposing suitable subtraction conditions. The numerical simulations have been carried out using the gauge field configurations produced by the European Twisted Mass Collaboration with $N_f = 2+1+1$ dynamical quarks at three values of the lattice spacing. Our result for the B-parameter of the chromomagnetic operator at the physical pion and kaon point is $B_{CMO}^{K \pi} = 0.273 ~ (69)$, while in the SU(3) chiral limit we obtain $B_{CMO} = 0.076 ~ (23)$. Our findings are significantly smaller than the model-dependent estimate $B_{CMO} \sim 1 - 4$, currently used in phenomenological analyses, and improve the uncertainty on this important phenomenological quantity.
\end{abstract}

\maketitle

\newpage

\section{Introduction}
\label{sec:intro}

At low energy with respect to the electroweak scale, the Standard Model (SM) and its possible New Physics (NP) extensions are described by an effective Hamiltonian in which the contribution of operators of dimension $d = 4+n$ are suppressed by $n$ powers of the high-energy (i.e.~the electroweak or NP) scale.
In the flavor changing $\Delta S = 1$ sector, the effective Hamiltonian contains four operators of dimension $d = 5$, two electromagnetic (EMO) and two chromomagnetic (CMO) operators.
Their contribution to the physical amplitudes is thus suppressed by only one power of the high-energy scale.
The $\Delta S = 1$, $d = 5$ effective Hamiltonian has the form
 \be
     H^{\Delta S = 1,\ d = 5}_{\rm eff} = \sum_{i = \pm} \left(C_\gamma^i \, Q^i_\gamma + C_g^i \, Q_g^i \right) + 
          {\rm h.c.} \ ,
     \label{eq:Heff}
 \ee
where $Q_{\gamma, g}^+$ ($Q_{\gamma, g}^-$) are the parity-even (-odd) EMO and CMO, respectively, defined as:
 \bea
      \label{eq:Qgamma_pm}
     Q_\gamma^\pm & = & {Q_d\,e\over 16 \pi^2} \left(\bar{s}_L \,\sigma^{\mu\nu}\, F_{\mu\nu} \, d_R
          \pm \bar{s}_R \,\sigma^{\mu\nu}\, F_{\mu\nu} \, d_L  \right) \ , \nonumber \\
     \label{eq:Qg_pm}
     Q_g^\pm & = & { g\over 16 \pi^2} \left(\bar{s}_L \,\sigma^{\mu\nu}\, G_{\mu\nu} \, d_R   
           \pm \bar{s}_R \,\sigma^{\mu\nu}\, G_{\mu\nu} \, d_L \right) \ ,
 \eea
with $q_{R, L} = \frac{1}{2}(1 \pm \gamma_5)\, q$ (for $q = s, d$).

\begin{figure}[htb!]
\centering
\includegraphics[scale=0.70]{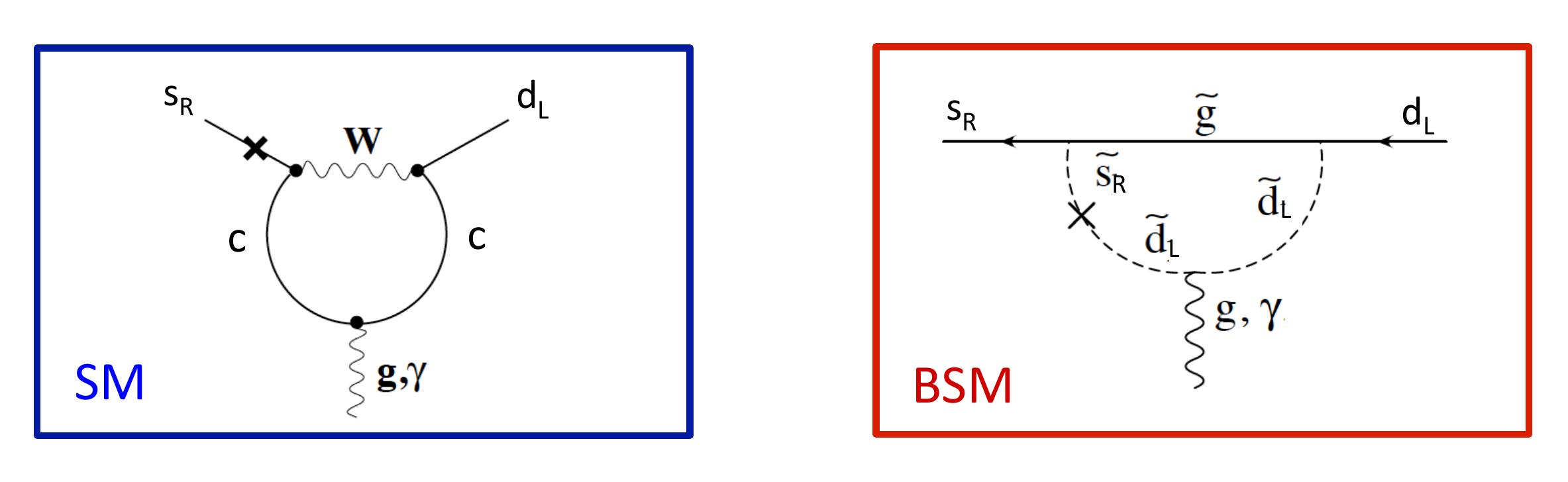}
\vspace{-1.0cm}
\caption{\it One-loop Feynman diagrams contributing at low energy to the effective magnetic interactions, in the SM (left) and beyond (right). In the latter case we have shown, for illustrative purposes, the case of SUSY models. The crosses denote a mass insertion.}
\label{fig:fd}
\end{figure}

In Fig.~\ref{fig:fd} we show two examples of Feynman diagrams generating, at low energy, the effective magnetic interactions in the SM and beyond.
For illustration of the NP contribution we have considered the case of SUSY models, in which the $\Delta S = 1$ transition occurs through the exchange of virtual squarks and gluinos and it is mediated by the strong interactions.

Note that at least one mass insertion is required in the diagrams, both in the SM and beyond, in order to induce the chirality flip described by the magnetic operators. 

A quick inspection of the diagrams of Fig.~\ref{fig:fd} shows that the Wilson coefficients of the magnetic operators in the SM and NP model are proportional to
 \be
     C_{\gamma, g}^{SM} \sim \frac{\alpha_W(M_W)}{M_W} \, \frac{m_s}{M_W} ~ , \qquad
     C_{\gamma, g}^{NP} \sim \frac{\alpha_s(M_{NP})}{M_{NP}} \, \delta_{LR} ~ ,
     \label{eq:cWilson}
 \ee
where $M_{NP}$ represents the typical NP scale, e.g.~the gluino mass in the SUSY case, and the factors $m_s / M_W$ and $\delta_{LR}$ are generated in the diagrams by the mass insertion.
In the SUSY case, for instance, $\delta_{LR}$ represents the off-diagonal matrix element of the squark mass matrix normalized to the average squark mass.
The transition rate is controlled in the SM by the weak coupling $\alpha_W(M_W)$.
This is not generally the case for NP models.
In the SUSY transition shown in Fig.~\ref{fig:fd}, for example, the process is mediated by the strong interactions. 
Therefore, the proportionality of $C_{\gamma,g}^{NP}$ in Eq.~(\ref{eq:cWilson}) to the strong coupling constant $\alpha_s(M_{NP})$, rather than to the weak coupling as in the SM, compensates in part for the stronger high-energy scale suppression ($M_{NP} > M_W$) in the NP model.
Thus, the magnetic interactions receive potentially large contributions from physics beyond the SM. 

It is also worth noting that the chirality flipping factor $m_s / M_W$, which appears in the Wilson coefficients of the magnetic operators in the SM, is of the same size as $\Lambda_{QCD} / M_W$, which represents the additional suppression factor of the coefficients of dimension-6 operators in the effective Hamiltonian.
For this reason, the role of the magnetic operators tends to be marginal in the SM, while it is potentially more relevant for the searches of NP.

The $K \to \pi$ matrix element of the EMO $Q_{\gamma}^+$, which is relevant for instance for the CP violating part of the rare $K_L \to \pi^0 \ell^+ \ell^-$ decays~\cite{Buras:1999da}, has been computed on the lattice both in the quenched~\cite{Becirevic:2000zi} and unquenched $N_f = 2$~\cite{Baum:2011rm} case.
Since the electromagnetic field strength tensor $F_{\mu\nu}$ factorizes out of the hadronic matrix element, the lattice computation only involves the quark bilinear operator $\bar s \,\sigma^{\mu\nu} d$, and it is relatively straightforward. 

Computing the hadronic matrix elements of the CMO $Q_g^\pm$ is, instead, by far more challenging. 
The main difficulty is represented by the complicated renormalization pattern of the operator, which also involves power divergent mixing with operators of lower dimensionality (see Ref.~\cite{Constantinou:2015ela} and Section \ref{sec:RC}).
The relevant matrix elements with an initial kaon involve one, two or three pions in the final states, and are of great phenomenological interest for various processes: the long distance contribution to $K^0 - \bar K^0$ mixing~\cite{He:1999bv}, $\Delta I = 1/2$, $K \to \pi \pi$ transitions and $\varepsilon^\prime / \varepsilon$~\cite{Buras:1999da}, CP violation in $K \to 3\,\pi$ decays~\cite{D'Ambrosio:1999jh}.
These matrix elements are parameterized in terms of suitably defined B-parameters:
 \bea
     \label{eq:me}
     \langle \pi^+ | Q_g^+ | K^+ \rangle & = & \frac{11}{32\pi^2}\, \frac{M_K^2 \, (p_K \cdot p_\pi)}{m_s + m_d} \, 
          B_{CMO}^{K\pi} \ , \\
    \label{eq:me_2pi}
     \langle \pi^+ \pi^- | Q_g^- | K^0 \rangle & = & i\, \frac{11}{32\pi^2}\, \frac{M_K^2 \, M_\pi^2}{f_\pi\,(m_s + m_d)} \, 
          B_{CMO}^{K2\pi} \ , \\
    \label{eq:me_3pi}
      \langle \pi^+ \pi^+ \pi^- | Q_g^+ | K^+ \rangle & = & - \frac{11}{16\pi^2}\, \frac{M_K^2 \, M_\pi^2}{f_\pi^2\,(m_s + 
          m_d)} \, B_{CMO}^{K3\pi} \ .
 \eea

At leading order (LO) in SU(3) chiral perturbation theory (ChPT), the CMO has a single representation in terms of pseudo-Goldstone boson fields~\cite{Bertolini:1994qk},
 \be
     \label{eq:chiral}
     Q_g^\pm = \frac{11}{256\pi^2}\, \frac{f_\pi^2 \, M_K^2}{m_s + m_d} \, B_{CMO} \, 
          \left[ U (D_\mu U^\dagger) (D^\mu U) \pm  (D_\mu U^\dagger) (D^\mu U) U^\dagger \right]_{23} \ ,
 \ee
where the low-energy constant $B_{CMO}$ is estimated to be of order 1 in the chiral quark model of Ref.~\cite{Bertolini:1994qk}.
Therefore, the three B-parameters of Eqs.~(\ref{eq:me}-\ref{eq:me_3pi}) are related by SU(3) chiral symmetry, which predicts at LO their equality: $B_{CMO}^{K\pi} = B_{CMO}^{K2\pi} = B_{CMO}^{K3\pi} = B_{CMO}$.
Such an equality is expected to be broken at higher orders in ChPT.

In this work we evaluate the B-parameter appearing in Eq.~(\ref{eq:me}) from the lattice QCD computation of the $\langle \pi | Q_g^+ | K \rangle$ matrix element.
We perform numerical simulations by employing the gauge configurations generated by the European Twisted Mass Collaboration (ETMC) with $N_f = 2 + 1 + 1$ dynamical quarks, which include in the sea, besides two light mass-degenerate quarks, also the strange and charm quarks with masses close to their physical values~\cite{Baron:2010bv,Baron:2011sf,Carrasco:2014cwa}.
The same gauge ensembles have been used in Ref.~\cite{Constantinou:2015ela} to determine non-perturbatively the power divergent mixing coefficients controlling the mixing of the CMO with operators of lower dimension. 
As for the mixing coefficients with operators of the same dimensionality and for the multiplicative renormalization constant (RC), we adopt the predictions of perturbation theory at one-loop obtained in Ref.~\cite{Constantinou:2015ela}.

Preliminary results for the B-parameter of the CMO have been presented in Ref.~\cite{Lubicz:2014qfa} and our final result at the physical pion and kaon point is
 \be
     \label{eq:BCMO}
     B_{CMO}^{K \pi} = 0.273 ~ (13) ~ (68) = 0.273 ~ (69) ~ ,
 \ee
where the first error comes from the numerical lattice simulations, while the second error accounts for the perturbative uncertainty in the one-loop determination of the multiplicative renormalization constant.
Our result~(\ref{eq:BCMO}) represents the first lattice QCD determination of a matrix element of the CMO.
In the SU(3) chiral limit we get $B_{CMO} = 0.076 ~ (23)$.
Our findings are significantly smaller than the model-dependent estimate $B_{CMO} \sim 1 - 4$ currently used in phenomenological analyses~\cite{Buras:1999da} and improve the uncertainty on this important phenomenological quantity.

The plan of the paper is as follows.
In section~\ref{sec:simulations} we describe the lattice setup and give the simulation details. 
In section~\ref{sec:RC} we recall the main results obtained in Ref.~\cite{Constantinou:2015ela} on the determination of the power-divergent mixing coefficients needed for the renormalization of the CMO. 
The mixing subtraction is evaluated in section~\ref{sec:correlators} and it is shown that the renormalized CMO correlator can be determined with a remarkable level of precision. 
The lattice data for the matrix elements of the renormalized CMO are presented in section~\ref{sec:CMO} and analyzed in terms of both SU(2) and SU(3) ChPT. 
 Finally, section~\ref{sec:conclusions} contains our conclusions.

\section{Simulation details}
\label{sec:simulations}

Our lattice setup is based on the gauge configurations generated by ETMC with $N_f = 2 + 1 + 1$ dynamical quarks~\cite{Baron:2010bv,Baron:2011sf}, adopted in Ref.~\cite{Carrasco:2014cwa} for the determination of the up, down, strange and charm quark masses, using the experimental value of the pion decay constant $f_\pi$ to set the lattice scale\footnote{Compared to Ref.~\cite{Carrasco:2014cwa} the number of independent gauge configurations adopted for the ensemble D15.48 has been increased to $90$ to improve the statistics.}. 

The gauge fields are simulated using the Iwasaki gluon action \cite{Iwasaki:1985we}, while sea quarks are implemented with the Wilson Twisted Mass Action at maximal twist \cite{Frezzotti:2000nk,Frezzotti:2003xj,Frezzotti:2003ni}. 
In order to avoid the mixing of strange and charm quarks induced by lattice artifacts in the unitary twisted mass formulation, we have adopted the mixed action setup described in Ref.~\cite{Frezzotti:2004wz}, where the valence strange quarks are regularized as Osterwalder-Seiler (OS) fermions \cite{Osterwalder:1977pc}, while the valence up and down quarks have the same action as the sea. 
The use of different lattice regularisations for the valence and sea quarks of the second generation preserves unitarity in the continuum limit and brings no complications for the operator renormalization pattern in mass-independent schemes, while producing only a modification of discretization effects.
Therefore, the uncertainty related to the use of a non-unitary action at finite lattice spacing is incorporated directly in the error due to discretization effects, which will be addressed in Section \ref{sec:CMO}.
Moreover, since we work with both valence and sea quarks at maximal twist, physical observables are guaranteed  to be automatically ${\cal{O}}(a)$-improved \cite{Frezzotti:2003ni,Frezzotti:2004wz}.

The details of the ETMC gauge ensembles with $N_f = 2+1+1$ dynamical quarks are collected in Table \ref{tab:simudetails}, where the number of the gauge configurations analyzed ($N_{cfg}$) corresponds to a separation of $20$ trajectories.
The QCD simulations are carried out at three different values of the inverse bare lattice coupling $\beta$, to allow for a controlled extrapolation to the continuum limit, and at different lattice volumes.
In this work the up- and down-quark masses are always taken to be degenerate ($m_u = m_d = m_{ud}$) and equal in the sea and valence sectors ($m_{ud}^{sea} = m_{ud}^{val} = m_{ud}$).
In the light sector we have simulated quark masses $m_{ud} = \mu_{ud} / Z_P$ in the range $3 ~ m_{ud}^{phys} \lesssim  m_{ud} \lesssim 12 ~ m_{ud}^{phys}$, where $m_{ud}^{phys}$ is the physical light-quark mass and $Z_m \equiv Z_P^{-1}$ is the mass renormalization constant (at maximal twist) determined in Ref.~\cite{Carrasco:2014cwa}.
In the strange sector we have used three values of the valence strange quark mass $m_s = \mu_s / Z_P$ in the range $0.7 ~ m_s^{phys} \lesssim  m_s \lesssim 1.2 ~ m_s^{phys}$, where $m_s^{phys}$ is the physical strange quark mass obtained in Ref.~\cite{Carrasco:2014cwa}.
The values of the lattice spacing are $a = \{ 0.0885\,(36), 0.0815\,(30), 0.0619\,(18)\}$ fm at $\beta = \{1.90, 1.95, 2.10\}$ respectively, the lattice volume goes from $\simeq 2$ to $\simeq 3$ fm and the pion masses, extrapolated to the continuum and infinite volume limits, range from $\simeq 210$ to $\simeq 450$ MeV (see Ref.~\cite{Carrasco:2014cwa} for further details).

\begin{table}[hbt!]
\centering
{\small
\begin{tabular}{||c|c|c|c|c|c|c|c||c|c||}
\hline
ensemble & $\beta$ & $V / a^4$ & $a\mu_{sea} = a\mu_{ud}$& $a\mu_\sigma$ & $a\mu_\delta$ & $N_{cfg}$ & $a\mu_s$ & 
$M_\pi(\mbox{MeV})$ & $M_K(\mbox{MeV})$ \\ \hline \hline
$A30.32$ & $1.90$ & $32^3 \times 64$ &$0.0030$ &$0.15$ &$0.19$ &$150$& $0.0145, 0.0185, 0.0225$ & $275$ & $577$ \\
$A40.32$ & & & $0.0040$ & & & $100$ & & $315$ & $588$ \\
$A50.32$ & & & $0.0050$ & & &  $150$ & & $350$ & $595$ \\ 
\hline \hline
$A40.24$ & $1.90$ & $24^3 \times 48 $ & $0.0040$ &$0.15$ & $0.19$& $150$ & $0.0145, 0.0185, 0.0225$ & $324$ & $594$ \\
$A60.24$ & & & $0.0060$ & & & $150$ & & $388$ & $610$ \\
$A80.24$ & & & $0.0080$ & & & $150$ & & $438$ & $624$ \\
$A100.24$ & & & $0.0100$ & & & $150$ & & $497$ & $650$ \\ 
\hline \hline
$B25.32$ & $1.95$ & $32^3 \times 64$ &$0.0025$&$0.135$ &$0.170$& $150$& $0.0141, 0.0180, 0.0219$ & $259$ & $553$ \\
$B35.32$ & & & $0.0035$  & & & $150$ & & $300$ & $562$ \\
$B55.32$ & & & $0.0055$  & & & $150$ & & $377$ & $587$ \\
$B75.32$ &  & & $0.0075$ & & & $80$ & & $437$ & $608$ \\ 
\hline \hline
$B85.24$ & $1.95$ & $24^3 \times 48 $ & $0.0085$ &$0.135$ &$0.170$ & $150$ & $0.0141, 0.0180, 0.0219$ & $463$ & $617$ \\
\hline \hline
$D15.48$ & $2.10$ & $48^3 \times 96$ &$0.0015$&$0.12$ &$0.1385 $& $90$& $0.0118, 0.0151, 0.0184$ & $224$ & $538$ \\ 
$D20.48$ & & & $0.0020$  &  &  & $100$ & & $255$ & $541$ \\
$D30.48$ & & & $0.0030$  &  &  & $100$ & & $310$ & $554$ \\
\hline
\end{tabular}
}
\vspace{0.25cm}
\caption{\it \small Parameters of the gauge ensembles and the values of the simulated sea and valence quark bare masses (in lattice units) used in this work (see Ref.~\cite{Carrasco:2014cwa} for details). The values of the kaon mass, $M_K$, do not correspond to a simulated strange bare quark mass shown in the $8^{\rm th}$ column, but to the renormalized strange mass interpolated at the physical value, $m_s^{phys}(\overline{MS}, 2~\rm{GeV}) = 99.6 (4.3)$ MeV, determined in Ref.~\cite{Carrasco:2014cwa}.}
\label{tab:simudetails}
\end{table}

Quark propagators are obtained using the multiple mass solver method \cite{Jegerlehner:1996pm, Jansen:2005kk}, which allows to invert the Dirac operator for several quark masses at a relatively low computational cost.   
The statistical accuracy of the meson correlators is significantly improved by using the ``one-end" stochastic method \cite{McNeile:2006bz}, which includes spatial stochastic sources at a single time slice chosen randomly.
Statistical errors are evaluated using the jackknife procedure.

\section{Renormalization of the chromomagnetic operator}
\label{sec:RC}

In this Section we briefly review the main results obtained in Ref.~\cite{Constantinou:2015ela} for the renormalization of the CMO, whose specific renormalization pattern depends on the details of the lattice regularization, i.e.~on the choice of the lattice action.
A detailed analysis of the implications of the discrete symmetries of the twisted-mass action was carried out in Ref.~\cite{Constantinou:2015ela}, showing that the renormalization of the CMO involves the mixing among 13  operators of equal or lower dimensionality\footnote{The operator mixing pattern considered in Ref.~\cite{Constantinou:2015ela} includes one operator ($O_6$), which is not independent from the other ones, and misses one five-dimensional operator~\cite{Bhattacharya:2015rsa}, which mixes with the CMO only at two loops in perturbation theory. The results presented in this work are therefore not affected.}, including also non gauge invariant operators vanishing by the equation of motion.

In the case of on-shell matrix elements the mixing simplifies, and the renormalized parity-even CMO can be written as~\cite{Constantinou:2015ela} 
 \be
    \label{eq:reno}
    \widehat{O}_{CM} = Z_{CM} \left[O_{CM} - \left( \frac{c_{13}}{a^2} + c_2 (\mu_s^2 + \mu_d^2) + c_3 \mu_s \mu_d \right) S - 
        \frac{c_{12}}{a} (\mu_s + \mu_d) P - c_4 O_4 \right] ,
 \ee
where $O_{CM} \equiv 16 \pi^2 Q_g^+ = g \bar{s} \sigma_{\mu\nu} G_{\mu\nu} d$, $S = \bar{s} d$,  $P = i \bar{s} \gamma_5 d$ and $O_4 = \Box(\bar{s} d)$ are bare operators, valence quarks are taken with the same value of the Wilson $r$-parameter, i.e.~$r_s = r_d$ (see Ref.~\cite{Constantinou:2015ela}), and $\mu_s$ ($\mu_d$) denotes the bare strange (light) quark mass.

Note that the quadratically divergent mixing of the CMO with the scalar density $S$ is common to any regularization, whereas the mixing with the pseudoscalar density $P$ (softened by the proportionality to the quark masses) is peculiar of twisted mass fermions, and it is a consequence of the non conservation of parity.
Moreover, in Eq.~(\ref{eq:reno}) the power divergent mixing coefficients $c_{12}$ and $c_{13}$ are scheme and renormalization scale independent \cite{Testa:1998ez}, while the multiplicative RC $Z_{CM}$ and the coefficients $c_i$ with $i = 2, 3, 4$ depend on both the scheme and the renormalization scale.

It is well known~\cite{Maiani:1991az} that the determination of power divergent coefficients, controlling the mixing with operators of lower dimension, cannot rely on perturbation theory. 
The reason is that potential non-analytic (in $g^2$) contributions to these coefficients, like those proportional to powers of $(1 / a) \exp(-1/(\beta_0 g^2)) \sim \Lambda_{QCD}$, do not appear in the perturbative expansion. 
Therefore, while for the present study the multiplicative renormalization factor $Z_{CM}$ and the coefficients $c_i$ with $i = 2, 3, 4$ have been evaluated in perturbation theory at one-loop, the coefficients $c_{12}$ and $c_{13}$  in Eq.~(\ref{eq:reno}) have been determined in a non-perturbative way by imposing two suitable subtraction conditions~\cite{Constantinou:2015ela}.
The first one is that the matrix element of the CMO between external kaon and pion at rest must vanish in the SU(3) chiral limit, namely
 \be
     \label{eq:c13}
     \frac{1}{Z_{CM}} \lim_{m_s, ~ m_d \to 0} ~ \langle \pi | \widehat O_{CM} | K \rangle = 
         \lim_{m_s, ~ m_d \to 0} ~ \langle \pi | O_{CM} - \frac{c_{13}}{a^2} S | K \rangle = 0 ~ ,
 \ee
from which the coefficient $c_{13}$ can be determined.
The second requirement is the vanishing of the parity violating matrix elements of the CMO up to terms of ${\cal{O}}(a)$, specifically
 \be
     \label{eq:c12}
     \frac{1}{Z_{CM}} \langle 0 | \widehat O_{CM} | K \rangle = \langle 0 | O_{CM} - \frac{c_{13}}{a^2} S -
         \frac{c_{12}}{a} (\mu_s + \mu_d) P | K \rangle = 0 ~ ,
 \ee
from which the coefficient $c_{12}$ can be calculated once the coefficient $c_{13}$ is determined from Eq.~(\ref{eq:c13}).

In Table~\ref{tab:coeff} we present the numerical results for the various mixing coefficients, obtained in Ref.~\cite{Constantinou:2015ela} at the three values of the inverse coupling $\beta$ given in Table \ref{tab:simudetails}.
The central values and the errors of the coefficient $c_{13}$ shown in the last column correspond to the averages and the spread of the two non-perturbative determinations corresponding to the choice ``LP" given in Table IV of Ref.~\cite{Constantinou:2015ela}.
For the mixing coefficient $c_{12}$ the uncertainty of the non-perturbative results has been found at the level of $\simeq 60 \%$~\cite{Constantinou:2015ela}, and therefore the $g^2$-dependence of the non-perturbative determination of $c_{12}$, shown in the penultimate column, can be safely neglected.
\begin{table}[htb!]
\begin{center}
\begin{tabular}{||c||c|c|c|c|c|c||c|c||}
\hline
$\beta$ & $Z_{CM}$ & $c_2$ & $c_3$ & $c_4$ & $c_{12}$ & $c_{13}$ & $c_{12}$ & $c_{13}$ \\ \hline
\multicolumn{1}{||c||}{} & \multicolumn{6}{|c||}{one-loop perturbation theory} & \multicolumn{2}{|c||}{non-perturbative} \\
\hline
1.90 & 1.781 & ~0.150 & ~0.0~ & ~0.0~ & 0.0854 & 0.962 & 0.035 (20) & 0.89713 (11) \\
1.95 & 1.752 & ~0.100 & ~0.0~ & ~0.0~ & 0.0832 & 0.937 & 0.035 (20) & 0.87629 (13) \\
2.10 & 1.677 &  -0.042 & ~0.0~ & ~0.0~ & 0.0772 & 0.870 & 0.035 (20) & 0.81676 ~(8) \\
\hline   
\end{tabular}
\end{center}
\caption{\it \small Values of the multiplicative renormalization factor $Z_{CM}$, in the $\overline{MS}$ scheme at the scale $\mu = 2$ GeV, and of the mixing coefficients $c_i$ in Eq.~(\protect\ref{eq:reno}), obtained in Ref.~\protect\cite{Constantinou:2015ela}. The results correspond to the three values of the inverse coupling $\beta$ given in Table \ref{tab:simudetails} using one-loop perturbation theory, except for $c_{12}$ and $c_{13}$ in the last two columns, which have been obtained non-perturbatively (see text). The perturbative results have been evaluated using the bare coupling $g_0^2 = 6 / \beta$ for the power divergent coefficients $c_{12}$ and $c_{13}$ and the boosted coupling $g_P^2 = g_0^2 /U_P$ for the other coefficients, where $U_P$ is the average plaquette equal to $\{0.575, 0.585, 0.614\}$ at $\beta = \{1.90, 1.95, 2.10\}$.}
\label{tab:coeff}
\end{table}

It can be seen that: ~ i) the power divergent coefficient $c_{13}$ has been determined non-perturbatively with a very high level of precision; ~ ii) the coefficients $c_3$ and $c_4$ vanish at one loop; ~ iii) the coefficient $c_2$ which starts at ${\cal O}(g^2)$ is rather small; and ~ iv) the multiplicative renormalization factor $Z_{CM}$ receives at one loop a sizable correction ($\sim 70 \%$).

In Table \ref{tab:coeff} we also provide the one-loop results for the power divergent coefficients $c_{12}$ and $c_{13}$.
For the latter the difference between the one-loop and the non-perturbative results is less than $10 \%$.
The bulk of the difference is compatible with being a correction of ${\cal{O}}(g^4)$. 
Thus, genuine non-perturbative contributions to $c_{13}$ are likely to be small, even though a firmer conclusion in this sense would require the calculation of $c_{13}$ at two loops at least. 

As far as the coefficient $c_{12}$ is concerned, its size is smaller by (at least) one order of magnitude with respect to $c_{13}$ both perturbatively and non-perturbatively (see Table \ref{tab:coeff}). 
In addition, the corresponding operator is proportional to the first power of the quark masses, with $a (m_s + m_d) \sim 0.02$ in our simulations.
For these reasons, the subtraction of the linear divergence in Eq.~(\ref{eq:reno}) has a numerically negligible impact on the determination of the CMO matrix elements (see next Section).

\section{Lattice QCD correlators}
\label{sec:correlators}

In order to evaluate the matrix elements of the renormalized CMO (\ref{eq:reno}), using the values of the mixing coefficients given in Table  \ref{tab:coeff}, we need to determine the matrix elements of three bare local operators: $O_{CM}$, $O_S \equiv S$ and $O_P \equiv P$.
For the scalar and pseudoscalar densities we adopt the local versions $S = \bar{s} d$ and $P = i \bar{s} \gamma_5 d$, respectively, while in the lattice version of the chromomagnetic operator $O_{CM}$ the gluon tensor $G_{\mu \nu}$ is replaced by its clover discretization $P_{\mu \nu}$, namely \cite{Gabrielli:1990us}
 \be
       \label{OCM_lattice}
       O_{CM} = g_0 ~ \overline{\psi}_s \sigma_{\mu \nu} P_{\mu \nu} \psi_d ~ , 
 \ee
where 
 \be 
    P_{\mu \nu}(x) \equiv \frac{1}{4 a^2} \sum_{j = 1}^4 \frac{1}{2 i g_0} \left[ P_j(x) - P_j^\dagger(x) \right]
    \label{Pmunu}
 \ee 
and the sum is over the four plaquettes $P_j(x)$ in the $\mu$-$\nu$ plane stemming from $x$ and taken in the counterclockwise sense.

The $K \to \pi$ matrix elements of the bare local operators $O_i$ ($ i = \{ CM, ~ S, ~ P \}$) are extracted from the large (Euclidean) time distance behavior of a convenient combination of 2- and 3-point correlation functions, which for both initial and final mesons at rest are defined as 
\bea
   \label{eq:C2}
    C^{\pi(K)}(t) & = & \frac{1}{L^3} \sum_{\vec{x}, \vec{z}} \langle P_{\pi(K)}(\vec{x}, t + t_z) P_{\pi(K)}^\dagger(\vec{z}, t_z) \rangle ~ , \\
   \label{eq:C3}
   C_i^{K\pi(\pi K)}(t, t^\prime) & = & \frac{1}{L^6} \sum_{\vec{x}, \vec{y}, \vec{z}} \langle P_{\pi(K)}(\vec{x}, t^\prime + t_z) 
                                                         O_i(\vec{y}, t + t_z) P_{K(\pi)}^\dagger(\vec{z}, t_z) \rangle ~ ,
\eea
where $t^\prime$ is the time distance between the source and the sink, $t$ is the time distance between the insertion of the operator $O_i$ and the source, while $P_K(x) = i \bar{s}(x) \gamma_5 u(x)$ and $P_\pi(x) = i \bar{d}(x) \gamma_5 u(x)$ are the local interpolating fields of the $K$ and $\pi$ mesons, respectively. 
The Wilson parameters $r$ of the two valence quarks in both initial and final mesons are always chosen to have opposite values, i.e.~$r_s = r_d = -r_u$, so that the squared meson masses differ from their continuum counterpart only by terms of order ${\cal{O}}(a^2 m \Lambda_{QCD})$ \cite{Frezzotti:2003ni}.

The statistical accuracy of the correlators (\ref{eq:C2}-\ref{eq:C3}) is significantly improved by using the all-to-all quark propagators evaluated with the so-called ``one-end" stochastic method \cite{McNeile:2006bz}, which includes spatial stochastic sources at a single time slice chosen randomly.
Statistical errors are evaluated using the jackknife procedure. 

At large time distances 2- and 3-point correlation functions behave as
 \bea
        \label{eq:C2_larget}
        C^{\pi(K)}(t) & ~ _{\overrightarrow{t \gg a}} ~ & \frac{|Z_{\pi(K)}|^2}{2M_{\pi(K)}} 
            \left[ e^{-M_{\pi(K)} t} + e^{-M_{\pi(K))} (T - t)} \right] , \\
        \label{eq:C3_larget}        
        C_i^{K\pi(\pi K)}(t, t^\prime) & ~ _{\overrightarrow{t \gg a, (t^\prime - t) \gg a}} ~ &  
            \frac{Z_{\pi(K)} Z_{K(\pi)}^*}{4M_\pi M_K} \langle K(\pi) |O_i | \pi(K) \rangle e^{-M_{K(\pi)} t} ~ e^{-M_{\pi(K)} (t^\prime - t)} ~ ,
 \eea
where $Z_{\pi(K)}$ is the matrix element $\langle 0 | P_{\pi(K)}(0) |\pi(K) \rangle$ and $M_{\pi(K)}$ is the mass of the $\pi(K)$ meson. 
Both quantities are determined adopting the fitting function (\ref{eq:C2_larget}) in the time interval $[t_{\rm min}, T/2]$, where $t_{\rm min}$ is the time distance at which the ground-state starts to dominate the 2-point correlator.
Explicitly we choose $t_{\rm min} / a = \{10, 12, 18 \}$ at $T / a = \{48, 64, 96 \}$ (cf.~Table \ref{tab:simudetails}) for both pion and kaon mesons.

The matrix elements $\langle K |O_i | \pi \rangle$ can be extracted from the time behavior of the following ratios
 \be
     R_i(t, t^\prime) =  s_i(t, t^\prime) \sqrt{4 M_\pi M_K \left| \frac{ C_i^{K \pi}(t, t^\prime) ~ C_i^{\pi K}(t, t^\prime)}
         {\widetilde{C}^\pi(t^\prime) ~ \widetilde{C}^K(t^\prime)} \right|} ~ ,
     \label{eq:ratio}
 \ee
where $s_i(t, t^\prime)$ is the sign of correlator $C_i^{K\pi}(t, t^\prime)$ and the correlation function $\widetilde{C}^{\pi(K)}(t)$ is given by
 \be
      \widetilde{C}^{\pi(K)}(t) \equiv \frac{1}{2}\left\{ C^{\pi(K)}(t) + \sqrt{\left[ C^{\pi(K)}(t) \right]^2 -
          \left[ C^{\pi(K)}(T/2) \right]^2} \right\} ~ ,
      \label{eq:C2_tilde}
 \ee
which at large time distances behave as
 \be
     \widetilde{C}^{\pi(K)}(t) ~ _{\overrightarrow{t \gg a}} ~ \frac{Z_{\pi(K)}}{2 M_{\pi(K)}} e^{-M_{\pi(K)} t} ~ ,
     \label{eq:C2_tilde_larget}
 \ee
i.e.~without the backward signal.
At large time distances one has
 \be
      R_i(t, t^\prime) ~ _{\overrightarrow{t \gg a, (t^\prime - t) \gg a}} ~ \langle K |O_i | \pi \rangle ~ ,
      \label{eq:ratio_larget}
 \ee
so that the bare matrix elements $\langle K | O_i | \pi \rangle$ can be calculated from the plateau of $R_i$ independently of the matrix elements $Z_{\pi}$ and $Z_K$ of the interpolating fields.
In order to minimize excited state effects, in what follows the source-sink separation is fixed to $t' = T/2$.
Therefore, the region of time distances, where both the initial and final ground states dominate leading to the plateau (\ref{eq:ratio_larget}), corresponds to $[t_{\rm min} , T/2 - t_{\rm min}]$. 
Such a time interval will be adopted to extract the CMO matrix elements.

\begin{figure}[htb!]
\includegraphics[scale=0.70]{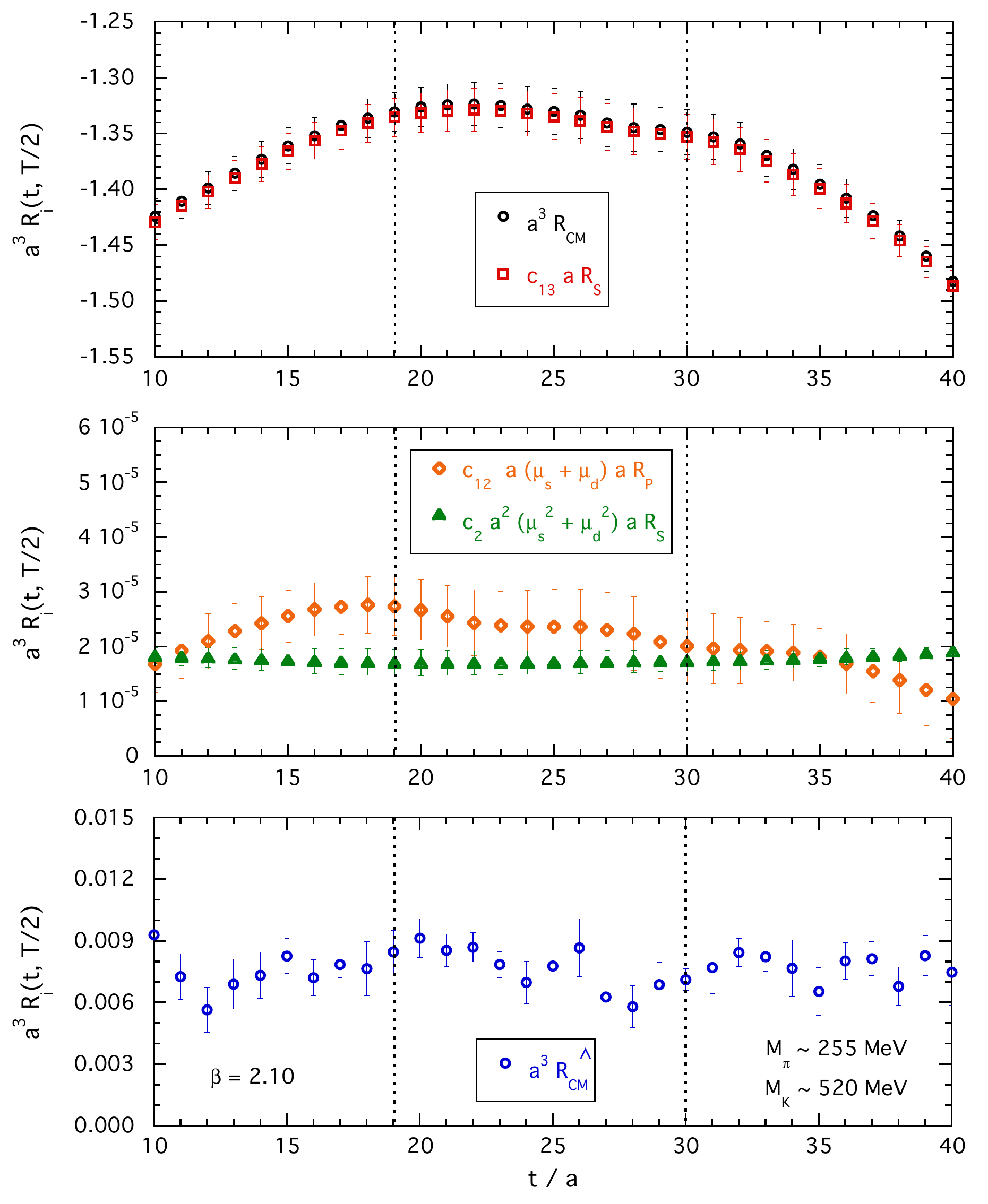}
\vspace{-0.5cm}\caption{\it \small Time dependence of the various terms contributing to Eq.~(\ref{eq:RCM}) in lattice units. Upper figure: $a^3 R_{CM}(t, T/2)$ and  $c_{13} a R_S(t, T/2)$. Middle figure: $c_{12} a (\mu_s + \mu_d) a R_P(t, T/2)$ and $c_2 a^2 (\mu_s^2 + \mu_d^2) a R_S(t, T/2)$. Bottom figure: the renormalized ratio $a^3 \widehat{R}_{CM}(t, T/2)$. The results refer to $\beta = 2.10$ and quark masses $a \mu_d = a \mu_{ud} = 0.0020$ and $a \mu_s = 0.0151$, corresponding to $M_\pi \simeq 255$ MeV and $M_K \simeq 520$ MeV (ensemble D20.48). The values adopted for the mixing coefficients $c_{12}$ and $c_{13}$ are the non-perturbative ones (see the last two columns of Table~\ref{tab:coeff}), while the value of the coupling entering $c_2$ is taken from boosted perturbation theory (see the third column of Table~\ref{tab:coeff}).  In the middle figure, the error of the term $c_{12} a (\mu_s + \mu_d) a R_P(t, T/2)$ is dominated by the large uncertainty of the non-perturbative determination of $c_{12}$ (see the penultimate column of Table~\ref{tab:coeff}).
For the ensemble D20.48 the time interval, where the plateau (\ref{eq:ratio_larget}) is expected to occur, is $[t_{\rm min} , T/2 - t_{\rm min}] / a = [18, 30]$ (see text) and it is indicated by the vertical dotted lines.}
\label{fig:ratios}
\end{figure}

The size of the various terms involved in the renormalization of the CMO (see Eq.~(\ref{eq:RCM}) below) and expressed in lattice units, namely $a^3 R_{CM}(t, T/2)$,  $c_{13} a R_S(t, T/2)$, $c_{12} a (\mu_s + \mu_d) a R_P(t, T/2)$ and $c_2 a^2 (\mu_s^2 + \mu_d^2) a R_S(t, T/2)$, can be inferred from Fig.~\ref{fig:ratios}, where the results refer to $\beta = 2.10$ and quark masses $a \mu_d = 0.0020$ and $a \mu_s = 0.0151$ (corresponding to $M_\pi \simeq 255$ MeV and $M_K \simeq 520$ MeV).
The values adopted for the mixing coefficients $c_{12}$ and $c_{13}$ are the non-perturbative ones, while the value of the coupling entering $c_2$ is taken from boosted perturbation theory (see Table~\ref{tab:coeff}).
It can be seen that the ratio $a^3 R_{CM}(t, T/2)$ and the one corresponding to the leading power divergence, $c_{13} a R_S(t, T/2)$, are almost equal, and the subtraction is at the level of $99.7 \%$. 
The other two terms, i.e.~$c_{12} a (\mu_s + \mu_d) a R_P(t, T/2)$ and $c_2 a^2 (\mu_s^2 + \mu_d^2) a R_S(t, T/2)$, are smaller by almost five orders of magnitude.   
The operators $\mu_s \mu_d S$ and $O_4 = \Box (\bar{s} d)$ do not contribute at one loop, since their mixing coefficients vanish (see Table \ref{tab:coeff}).

In the lower panel of Fig.~\ref{fig:ratios} we show the renormalized CM ratio $\widehat{R}_{CM}(t, T/2)$, given by
\bea
     \widehat{R}_{CM}(t, T/2) & = & Z_{CM} \left\{ R_{CM}(t, T/2) - \frac{c_{13}}{a^2} R_S(t, T/2) 
                                                       \right. \nonumber \\[2mm] 
                                             & - & \left. \frac{c_{12}}{a} (\mu_s + \mu_d) R_P(t, T/2) - 
                                                      c_2 (\mu_s^2 + \mu_d^2)  R_S(t, T/2) \right\} ~ .
     \label{eq:RCM}
 \eea
Despite being the outcome of a large numerical subtraction, due almost totally to the power divergent term $(c_{13} / a^2) R_S$, the results for $\widehat{R}_{CM}(t, T/2)$ are clearly different from zero and exhibit plateaux, from which the renormalized CMO matrix element $\langle K | \widehat{O}_{CM} | \pi \rangle$ can be determined quite precisely (see next Section).
In this respect we stress that the high level of precision achieved in the non-perturbative determination of $c_{13}$ (see the last column of Table \ref{tab:coeff}) plays a crucial role.

Notice that the renormalized CMO ratio $a^3 \widehat{R}_{CM}(t, T/2)$ is two orders of magnitude larger than the mixing term $c_2 a^2 (\mu_s^2 + \mu_d^2) a R_S(t, T/2)$.
Since $c_2$ (as well as $c_3$ and $c_4$ in Eq.~(\ref{eq:reno})) is known only at one loop in perturbation theory, higher-order corrections and non-perturbative effects might contribute to the the renormalized CMO ratio away from the SU(3) chiral limit. 
However, the smallness of the mixing term $c_2 a^2 (\mu_s^2 + \mu_d^2) a R_S(t, T/2)$ indicates that higher orders and non-perturbative effects are not expected to play a significant role in the determination of the matrix elements of the renormalized CMO within the present statistical uncertainties.

\section{Matrix elements of the chromomagnetic operator}
\label{sec:CMO}

By having defined the properly renormalized CMO, the matrix elements $\langle K | \widehat{O}_{CM} | \pi \rangle$ can be extracted from the plateaux of the ratio $\widehat{R}_{CM}(t, T/2)$ in the expected time interval $[t_{\rm min}, T/2 - t_{\rm min}]$, where $t_{\rm min}$ is the time distance at which excited states have decayed sufficiently from both the source and the sink, namely $t_{\rm min} = \{ 10, 12, 18 \}$ at $\beta = \{ 1.90, 1.95, 2.10 \} $. 
As described in Ref.~\cite{Frezzotti:2003ni}, the resulting matrix element $\langle K | \widehat{O}_{CM} | \pi \rangle$ is automatically ${\cal{O}}(a)$-improved.

The B-parameter $B_{CMO}^{K \pi}$, defined in Eq.~(\ref{eq:me}), is easily obtained as
 \be
     B_{CMO}^{K \pi} =  \frac{32 \pi^2}{11} \frac{m_s + m_d}{M_K^2} 
                                    \frac{\langle K | \widehat{O}_{CM} | \pi \rangle}{M_K M_\pi}  ~ .  
     \label{eq:BKPi}
 \ee
Note that the evaluation of the B-parameter (\ref{eq:BKPi}) does not require the knowledge of the lattice spacing, while it involves the mass RC $Z_m$, which in our maximally twisted-mass setup is given by $Z_m = 1/ Z_P$, where $Z_P$ is the RC of the pseudoscalar density.
For the latter we adopt the RI'-MOM results obtained in Ref.~\cite{Carrasco:2014cwa} using the two methods M1 and M2, which differ by $O(a^2)$ effects. 

Besides $B_{CMO}^{K \pi}$ we have calculated two further B-parameters.
They correspond to the transitions induced by the CMO in which either the mass $m_s$ of the strange valence quark is taken to be equal to the light-quark mass $m_d$ or the mass $m_d$ of the light valence quark is taken to be equal to the strange quark mass $m_s$.
In both cases the spectator valence quark, which does not participate in the transition, is always a up-quark with mass $m_{ud}$ (see Eqs.~(\ref{eq:C2}-\ref{eq:C3})).
We will refer to the above two transitions as the $\pi \pi$ and $K K$ channels, respectively.
Explicitly one has
 \bea
     \label{eq:BPiPi}
     B_{CMO}^{\pi \pi} =  \frac{32 \pi^2}{11} \frac{2 m_d}{M_\pi^2} 
                                      \frac{\langle \pi | \widehat{O}_{CM} | \pi \rangle}{M_\pi^2}  ~ , \\
     \label{eq:BKK}
     B_{CMO}^{K K} =  \frac{32 \pi^2}{11} \frac{m_s + m_d}{M_K^2} 
                                    \frac{\langle K | \widehat{O}_{CM} | K \rangle}{M_K^2}  ~ , 
 \eea
where $\langle \pi | \widehat{O}_{CM} | \pi \rangle \equiv [ \langle K | \widehat{O}_{CM} | \pi \rangle]_{m_s = m_d = m_{ud}}$ and $\langle K | \widehat{O}_{CM} | K \rangle \equiv [ \langle K | \widehat{O}_{CM} | \pi \rangle]_{m_d = m_s}$.
The $\pi \pi$ and $K K$ channels do not correspond to any physical process, but the set of the three quantities $B_{CMO}^{\pi \pi}$, $B_{CMO}^{K \pi}$ and $B_{CMO}^{K K}$ can be analyzed in terms of SU(3) ChPT\footnote{In Eq.~(\ref{eq:BKK}) the B-parameter $B_{CMO}^{K K}$ is defined in such a way as to guarantee that at LO in SU(3) ChPT the three B-parameters $B_{CMO}^{\pi\pi}$, $B_{CMO}^{K\pi}$ and $B_{CMO}^{KK}$ are always normalized by the same quark condensate in the chiral limit.}.

\begin{figure}[htb!]
\centering
\includegraphics[scale=0.70]{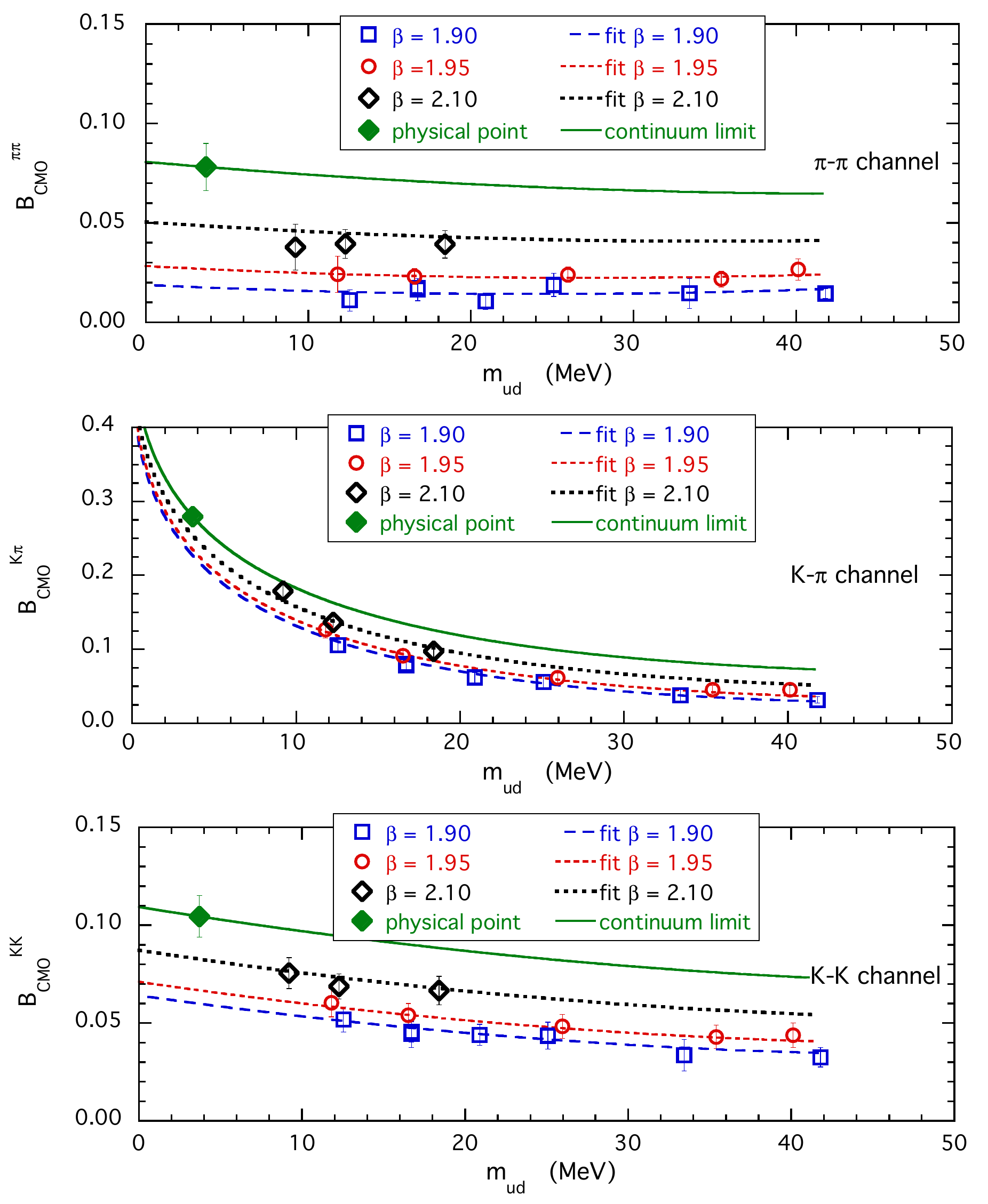}
\vspace{-0.5cm}
\caption{\it \small Values of the B-parameters $B_{CMO}^{\pi\pi}$ (upper), $B_{CMO}^{K\pi}$ (middle) and $B_{CMO}^{KK}$ (bottom), defined in Eqs.~(\ref{eq:BPiPi}), (\ref{eq:BKPi}) and (\ref{eq:BKK}), respectively, as a function of the renormalized light-quark mass $m_{ud}$ in the $\overline{\rm MS}(2~\mbox{\rm GeV})$ scheme. The lattice data for $B_{CMO}^{K \pi}$ and $B_{CMO}^{K K}$ have been smoothly interpolated to the physical value of the strange quark mass $m_s^{phys} = 99.6 (4.3)$ MeV~\cite{Carrasco:2014cwa}. Note the different scales in the three panels. The dashed, short-dashed and dotted lines correspond to the results of the SU(3)-inspired fit (\ref{eq:SU3fit}) for the three values of the lattice spacing, while the solid lines are the result in the continuum limit. The diamonds represent the values of the B-parameters at the physical pion mass and in the continuum limit.}
\label{fig:BCMO3}
\end{figure}

In Fig.~\ref{fig:BCMO3} we show the results for the B-parameters $B_{CMO}^{\pi\pi}$, $B_{CMO}^{K\pi}$ and $B_{CMO}^{KK}$ as a function of the renormalized light-quark mass $m_{ud}$ for the ETMC ensembles of Table \ref{tab:simudetails} in the $\overline{\rm MS}(2~\mbox{\rm GeV})$ scheme.
Since we have simulated three values of the strange quark mass around its physical value (see Table \ref{tab:simudetails}), the results for $B_{CMO}^{K \pi}$ and $B_{CMO}^{K K}$, shown in Fig.~\ref{fig:BCMO3}, correspond to a smooth interpolation at $m_s = m_s^{phys} = 99.6 (4.3)$ MeV, determined in Ref.~\cite{Carrasco:2014cwa}.
It can be seen that, for all the three channels, the results for the B-parameters exhibit controllable discretization effects, which are the beneficial consequence of the subtraction of the power-divergent mixings (thanks to the use of maximally
twisted Wilson quarks the subtracted CMO matrix elements vanish in the chiral limit 
even at finite lattice spacing).
Moreover, the impact of finite volume effects can be estimated by comparing the results corresponding to the ensembles A40.24 and A40.32, which share common values of the pion mass and the lattice spacing, and differ only by lattice size $L$.
No significant effects are visible within the statistical errors.
 
As described in section~\ref{sec:intro}, the ChPT prediction at LO is that all three B-parameters should coincide and be independent of the light-quark mass.
The latter feature is approximately fulfilled only in the case of $\pi \to \pi$ channel (see upper panel in Fig.~\ref{fig:BCMO3}).
Chiral corrections beyond the LO are clearly visible for both the $K \pi$ and $K K$ channels (see middle and bottom panels in Fig.~\ref{fig:BCMO3}), since they exhibit a remarkable dependence on the light-quark mass and deviate strongly from the results corresponding to the $\pi \pi$ channel.

In order to extrapolate to the physical pion mass we need to take into account at least NLO effects, which however are not known analytically.
We observe, in this respect, that the chiral corrections for the CMO matrix element contain also powers of $p_K \cdot p_\pi$.
This means that for the B-parameter $B_{CMO}^{K \pi}$ NLO terms proportional not only to $M_K^2$ and $M_\pi^2$, but also to $M_K M_\pi$ should be considered. 
Note that the quantity $M_K M_\pi \propto m_{ud}^{1/2}$ is non-analytic in the light-quark mass. 
In the $\pi \pi$ and $K K$ channels such a non-analytic term is not expected.

Thus, we introduce the variables
 \bea
     \xi_\pi & \equiv & \frac{2 B_0 m_{ud}}{(4 \pi f_0)^2} ~ , \nonumber \\[2mm]
     \xi_K & \equiv & \frac{B_0 (m_s + m_{ud})}{(4 \pi f_0)^2} ~ ,
  \eea
where $B_0$ and $f_0$ are LO low-energy constants (LECs) of the SU(3) ChPT~\cite{FLAG}, and we consider for the generic channel $i \to j$ with $i(j) = K, \pi$
the following SU(3)-inspired Ansatz
 \bea
      B_{CMO}^{i j} & = & B_{CMO} \left[ 1 + \alpha_1 \xi_\pi + \beta_1 (\xi_i + \xi_j) + \gamma_1 (\sqrt{\xi_i} - \sqrt{\xi_j})^2 
                                       \right. \nonumber \\
                             & + & \left. \alpha_2 \xi_\pi^2 + \beta_2 (\xi_i + \xi_j)^2 + \beta_2^\prime (\xi_i - \xi_j)^2 
                                       + \gamma_2 (\sqrt{\xi_i} - \sqrt{\xi_j})^4 \right] \nonumber \\
                             & + & a^2 \left[ D_0 + D_1 (\xi_i + \xi_j) \right] ~ ,
      \label{eq:SU3fit}
 \eea
where $B_{CMO}$ is the LO LEC appearing in Eq.~(\ref{eq:chiral}) (i.e., the SU(3) chiral limit of the B-parameters), while the parameters $\alpha_1$, $\beta_1$ and $\gamma_1$ play the role of NLO LECs, and $\alpha_2$, $\beta_2$, $\beta_2^\prime$ and $\gamma_2$ are NNLO LECs.
In Eq.~(\ref{eq:SU3fit}) the two terms proportional to $\xi_\pi$ and $\xi_\pi^2$ are due to the dependence on the mass of the u-quark (not involved in the transition) in common for all the channels.
Note that the NNLO term proportional to $(\sqrt{\xi_i} - \sqrt{\xi_j})^4$ generates in the $K \pi$ channel non-analytic terms proportional to $m_{ud}^{1/2}$ and $m_{ud}^{3/2}$.

According to SU(3) ChPT the LECs are independent of the light and strange quark masses, and therefore the Ansatz (\ref{eq:SU3fit}) can be applied to the combined analysis of the $\pi \pi$, $K \pi$ and $K K$ channels.
Taking into account that we have simulated three different values of the strange quark mass, the total number of lattice data is 105 (15 points for the $\pi \pi$ channel and 45 points for both $K \pi$ and $K K$ channells) and the number of fitting parameters is 10, whose values have been determined using a $\chi^2$-minimization procedure based on an uncorrelated $\chi^2$.
The results of the SU(3)-inspired fit (\ref{eq:SU3fit}) are shown in Fig.~\ref{fig:BCMO3}, where it can be seen that the quality of the fit, corresponding to $\chi^2 / \mbox{\rm d.o.f.} \simeq 0.5$, is remarkably good\footnote{In Eq.~(\ref{eq:SU3fit}) an additional NNLO term proportional to $(\xi_i + \xi_j) (\sqrt{\xi_i} - \sqrt{\xi_j})^2$ can be considered. We have checked that the impact of its inclusion is almost negligible. Moreover, the inclusion of additional discretization terms proportional either to $a^2 \xi_\pi$ or to $a^2 (\sqrt{\xi_i} - \sqrt{\xi_j})^2$ produces no significant effect within the errors.}.

After chiral and continuum extrapolations we get at the physical point: $B_{CMO}^{\pi \pi}|_{\rm phys} = 0.078 ~ (12)$, $B_{CMO}^{K \pi}|_{\rm phys} = 0.279 ~ (10)$ and $B_{CMO}^{K K}|_{\rm phys} = 0.105 ~(11)$, where the errors are statistical only.
The value of the LO LEC $B_{CMO}$, i.e. the SU(3) chiral limit  of the B-parameters, is close to the physical result for the $\pi \pi$ channel, namely $B_{CMO} = 0.076 ~ (14)$. 
Our results imply that in the $K \pi$ channel the impact of chiral orders higher than the LO corresponds to a strong enhancement factor equal to $\approx 4$.
Therefore, since higher orders in ChPT contribute differently in $K \to \pi$, $K \to \pi \pi$ and $K \to \pi \pi \pi$ transitions, the corresponding B-parameters are no more simply related to each other beyond the LO. 

The exclusion of all the NNLO terms in Eq.~(\ref{eq:SU3fit}) (i.e., putting $\alpha_2 = \beta_2 = \beta_2^\prime = \gamma_2 = 0$) leads to a lower quality fit having $\chi^2 / \mbox{\rm d.o.f.} \simeq 1.5$.
This is not surprising, since for a physical strange quark the impact of NNLO terms is expected to be non negligible.    

The result for $B_{CMO}^{K \pi}|_{\rm phys}$ is sensitive to the presence of the structures proportional to $(\sqrt{\xi_i} - \sqrt{\xi_j})^2$ and $(\sqrt{\xi_i} - \sqrt{\xi_j})^4$.
Putting $\gamma_2 = 0$ in Eq.~(\ref{eq:SU3fit}) we obtain $B_{CMO}^{K \pi}|_{\rm phys} = 0.267 ~ (10)$ with $\chi^2 / \mbox{\rm d.o.f.} \simeq 0.6$.
If all the non-analytic terms are neglected (i.e., $\gamma_1 = \gamma_2 = 0$), the quality of the corresponding fit deteriorates significantly ($\chi^2 / \mbox{\rm d.o.f.} \simeq 2.8$).

Adopting the SU(3)-inspired fit (\ref{eq:SU3fit}) and averaging the results of the different fits using Eq.~(28) of Ref.~\cite{Carrasco:2014cwa}, our result for the B-parameter $B_{CMO}^{K \pi}|_{\rm phys}$ is
 \be
     \label{eq:BCMO_SU3}
     B_{CMO}^{K \pi}|_{\rm phys} = 0.272 ~ (10)_{stat+fit} ~ (6)_{chir} ~ (6)_{disc} ~ (3)_{Z_P} = 0.272 ~ (13) ~ , 
 \ee
where
\begin{itemize}

\item $()_{stat+fit}$ indicates the uncertainty induced by both the statistical errors and the fitting procedure itself;

\item $()_{chir}$ corresponds to the uncertainty related to the chiral extrapolation, obtained using the results corresponding to the inclusion ($\gamma_2 \neq 0$) or the exclusion ($\gamma_2 = 0$) of the NNLO non-analytic term in Eq.~(\ref{eq:SU3fit});

\item $()_{disc}$ is the uncertainty related to discretization effects estimated by adding a term proportional to $a^4$ without any prior;

\item $()_{Z_P}$ is the error induced by the use of the two methods M1 and M2 to obtain the mass RC $Z_m = 1 / Z_P$ in Ref.~\cite{Carrasco:2014cwa}.

\end{itemize}

\begin{figure}[htb!]
\centering
\includegraphics[scale=0.70]{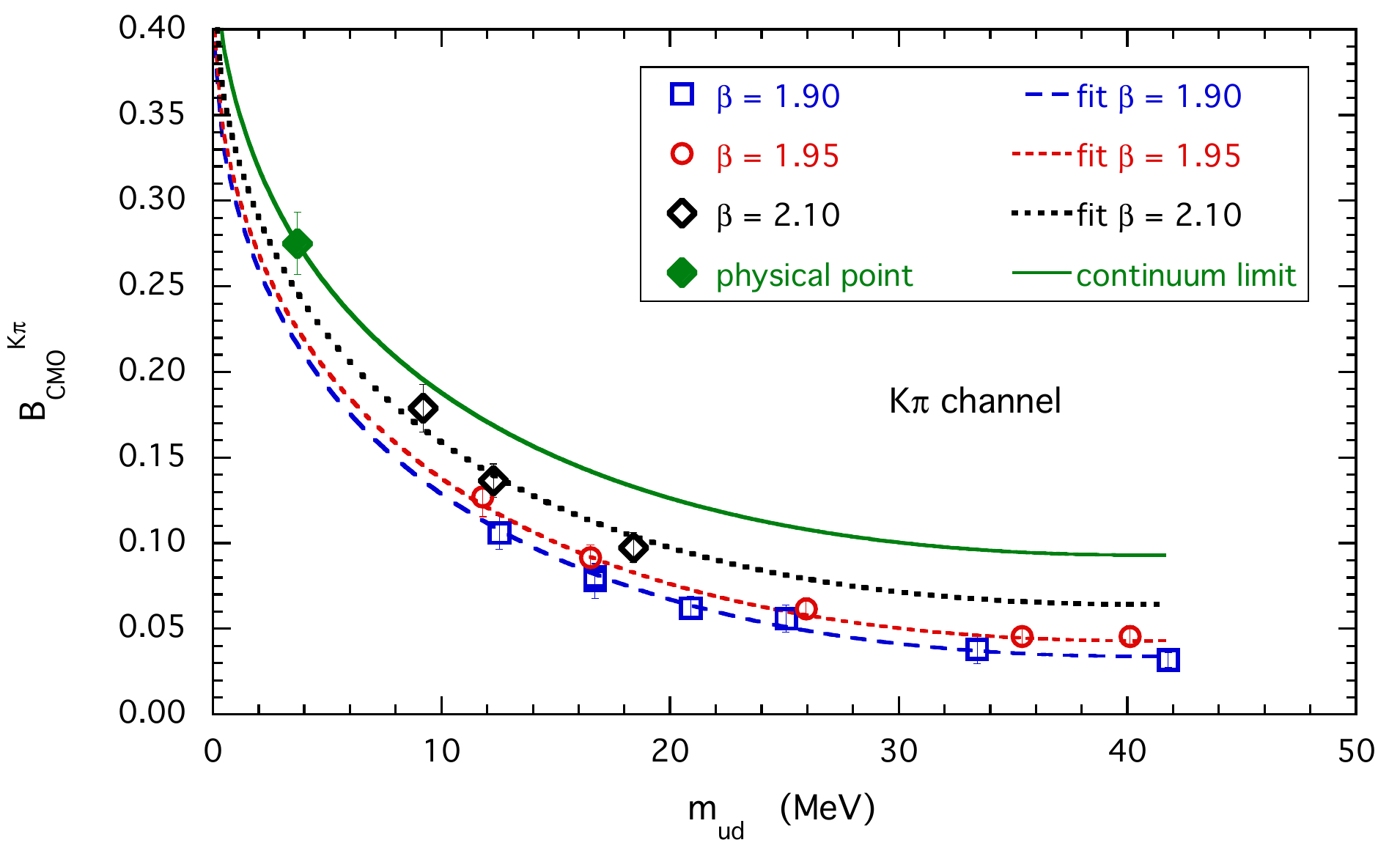}
\vspace{-0.5cm}
\caption{\it \small Values of the B-parameter $B_{CMO}^{K\pi}$ as a function of the renormalized light-quark mass $m_{ud}$ in the $\overline{\rm MS}(2~\mbox{\rm GeV})$ scheme. The lattice data have been smoothly interpolated to the physical value of the strange quark mass $m_s^{phys} = 99.6 (4.3)$ MeV~\cite{Carrasco:2014cwa}. The dashed, short-dashed and dotted lines correspond to the results of the SU(2)-inspired fit (\ref{eq:SU2fit}) assuming $\Delta = 0$ for the three values of the lattice spacing, while the solid line is the result in the continuum limit. The diamonds represent the values of the B-parameters at the physical pion mass and in the continuum limit.}
\label{fig:BCMO2}
\end{figure}

As a further check, we have also analyzed separately the data for the $K \pi$ channel (see also Ref.~\cite{Lubicz:2014qfa}) obtained after a smooth interpolation at the physical value of the strange quark mass $m_s^{phys} = 99.6 (4.3)$ MeV~\cite{Carrasco:2014cwa}.
This allows us to adopt the following SU(2)-inspired Ansatz
 \be
     B_{CMO}^{K \pi} =  \alpha + \beta m_{ud} + \gamma m_{ud}^{1/2} + \Delta + D a^2
     \label{eq:SU2fit}
\ee
where the parameters $\alpha$, $\beta$ and $\gamma$ play the role of $SU(2)$ LECs, while the function $\Delta$ includes chiral corrections beyond the NLO. 
Then, we have performed fits of the 15 lattice points of the $K \pi$ channel (interpolated at $m_s = m_s^{phys}$) adopting three choices for $\Delta$, namely $\Delta = 0$, $\Delta \propto m_{ud}^{3/2}$ and $\Delta \propto m_{ud}^2$.
For all the three choices we obtain a good description of the lattice data ($\chi^2 / \mbox{\rm d.o.f.} \simeq 0.4, 0.2$ and $0.2$, respectively).
The extrapolation to the physical pion point, in the three cases, yields: $B_{CMO}^{K \pi}|_{\rm phys} = 0.275 ~ (18)$ ($\Delta = 0$), $B_{CMO}^{K \pi}|_{\rm phys} = 0.340 ~ (44)$ ($\Delta \propto m_{ud}^{3/2}$) and $B_{CMO}^{K \pi}|_{\rm phys} = 0.327 ~ (51)$ ($\Delta \propto m_{ud}^2$).
The results of the fit (\ref{eq:SU2fit}) assuming $\Delta = 0$ are shown in Fig.~\ref{fig:BCMO2}.

Adopting the SU(2)-inspired fit (\ref{eq:SU2fit}) and averaging the results of the different fits using Eq.~(28) of Ref.~\cite{Carrasco:2014cwa}, we obtain
 \be
     \label{eq:BCMO_SU2}
     B_{CMO}^{K \pi}|_{\rm phys} = 0.306 ~ (37)_{stat+fit} ~ (44)_{chir} ~ (16)_{disc} ~ (7)_{Z_P} = 0.306 ~ (60) ~ , 
 \ee
which is consistent with the SU(3) result (\ref{eq:BCMO_SU3}) though with much larger statistical and systematic uncertainties.

Thus, performing a weighted average of the SU(2) and SU(3) results we quote for $B_{CMO}^{K \pi}|_{\rm phys}$ the final value
 \be
    B_{CMO}^{K \pi}|_{\rm phys} = 0.273 ~ (13) ~ (68)_{PT} = 0.273 ~ (69)  ~ , 
    \label{eq:final}
 \ee
where the second error accounts for the perturbative uncertainty in the one-loop determination of the multiplicative RC $Z_{CM}$ (see Table~\ref{tab:coeff}) and it has been estimated to be $\simeq 25 \%$ relying on the difference between the values of $Z_{CM}$ obtained with and without boosted perturbation theory.
As it can be seen, this error represents the largest source of uncertainty in the determination of the $B_{CMO}^{K \pi}$ parameter.

Our result~(\ref{eq:final}) represents the first lattice QCD determination of a matrix element of the CMO.
In the SU(3) chiral limit we obtain $B_{CMO} = 0.076 ~ (14) ~ (18)_{PT} = 0.076 ~ (23)$.
Our findings are significantly smaller than the model-dependent estimate $B_{CMO} \sim 1 - 4$ currently adopted in phenomenological analyses~\cite{Buras:1999da}.
The comparison indicates also that the uncertainty on this important phenomenological quantity has been significantly reduced by lattice QCD.
A further drastic improvement in the precision can be achieved by removing the uncertainty due to the use of one-loop perturbation theory for estimating the multiplicative RC $Z_{CM}$, which is by far the dominating source of uncertainty in our results.

\section{Conclusions}
\label{sec:conclusions}

We have presented the results of the first lattice QCD calculation of the $K \to \pi$ matrix elements of the chromomagnetic operator $O_{CM} = g\, \bar s\, \sigma_{\mu\nu} G_{\mu\nu} d$, which appears in the effective Hamiltonian describing $\Delta S = 1$ transitions in and beyond the Standard Model. 

Having dimension 5, the chromomagnetic operator is characterized by a rich pattern of mixing with operators of equal and lower dimensionality. 
 The power divergent coefficients controlling the mixing with operators of lower dimension were determined non-perturbatively in Ref.~\cite{Constantinou:2015ela}, while the multiplicative renormalization factor as well as the mixing coefficients with the operators of equal dimension have been computed at one-loop in perturbation theory~\cite{Constantinou:2015ela}.
The precision achieved in the non-perturbative evaluation of the mixing with the scalar density leads to an extraction of the matrix element of the renormalized chromomagnetic operator with good accuracy.

The numerical simulations have been carried out using the gauge field configurations produced by ETMC with $N_f = 2+1+1$ dynamical quarks at three values of the lattice spacing. 
Our result for the B-parameter of the chromomagnetic operator at the physical pion and kaon point is $B_{CMO}^{K \pi} = 0.273 ~ (69)$, 
while in the SU(3) chiral limit we get $B_{CMO} = 0.076 ~ (23)$.
Our findings are significantly smaller than the model-dependent estimate $B_{CMO} \sim 1 - 4$, currently used in phenomenological analyses, and improve the uncertainty on this important phenomenological quantity.

\section*{Acknowledgments}

We warmly thank Jean-Marc Gerard for pointing out the proper chiral normalization of the $B_{CMO}^{KK}$ parameter.
We gratefully acknowledge the CPU time provided by PRACE under the project PRA027 ``QCD Simulations for Flavor Physics in the Standard Model and Beyond'' on the JUGENE BG/P system at JSC (Germany) and by the agreement between INFN and CINECA under the specific initiative INFN-LQCD123 on the Fermi BG/Q system at CINECA (Italy).
V.L., G.M. and S.S.~thank MIUR (Italy) for partial support under the contract PRIN 2015P5SBHT.


\begin{thebibliography}{99}

\bibitem{Buras:1999da}
  A.~J.~Buras, G.~Colangelo, G.~Isidori, A.~Romanino and L.~Silvestrini,
  Nucl.\ Phys.\ B {\bf 566} (2000) 3
  [hep-ph/9908371].

\bibitem{Becirevic:2000zi}
  D.~Becirevic {\it et al.} [SPQcdR Collaboration],
  Phys.\ Lett.\ B {\bf 501} (2001) 98
  [hep-ph/0010349].

\bibitem{Baum:2011rm}
  I.~Baum, V.~Lubicz, G.~Martinelli, L.~Orifici and S.~Simula,
  Phys.\ Rev.\ D {\bf 84} (2011) 074503
  [arXiv:1108.1021 [hep-lat]].

\bibitem{Constantinou:2015ela}
  M.~Constantinou, M.~Costa, R.~Frezzotti, V.~Lubicz, G.~Martinelli, D.~Meloni, H.~Panagopoulos and S.~Simula,
  Phys.\ Rev.\ D {\bf 92} (2015) no.3,  034505
  [arXiv:1506.00361 [hep-lat]].

\bibitem{He:1999bv}
  X.~G.~He, H.~Murayama, S.~Pakvasa and G.~Valencia,
  Phys.\ Rev.\ D {\bf 61} (2000) 071701
  [hep-ph/9909562].

\bibitem{D'Ambrosio:1999jh}
  G.~D'Ambrosio, G.~Isidori and G.~Martinelli,
  Phys.\ Lett.\ B {\bf 480} (2000) 164
  [hep-ph/9911522].

\bibitem{Bertolini:1994qk}
  S.~Bertolini, J.~O.~Eeg and M.~Fabbrichesi,
  Nucl.\ Phys.\ B {\bf 449} (1995) 197
  [hep-ph/9409437].

\bibitem{Baron:2010bv}
  R.~Baron {\it et al.} [ETM Collaboration],
  JHEP {\bf 1006} (2010) 111
  [arXiv:1004.5284 [hep-lat]].

\bibitem{Baron:2011sf}
  R.~Baron {\it et al.}  [ETM Collaboration],
  PoS LATTICE {\bf 2010} (2010) 123
  [arXiv:1101.0518 [hep-lat]].

\bibitem{Carrasco:2014cwa}
  N.~Carrasco {\it et al.}  [ETM Coll.],
  Nucl.\ Phys.\ B {\bf 887} (2014) 19, 
  [arXiv:1403.4504 [hep-lat]].

\bibitem{Lubicz:2014qfa}
  M.~Constantinou, M.~Costa, R.~Frezzotti, V.~Lubicz, G.~Martinelli, D.~Meloni, H.~Panagopoulos and S.~Simula,
  PoS LATTICE {\bf 2014} (2014) 390
  [arXiv:1412.1351 [hep-lat]].

\bibitem{Iwasaki:1985we}
  Y.~Iwasaki,
  Nucl.\ Phys.\ B {\bf 258} (1985) 141.

\bibitem{Frezzotti:2000nk}
  R.~Frezzotti, P.~A.~Grassi, S.~Sint and P.~Weisz,
  JHEP {\bf 0108} (2001) 058,
  [hep-lat/0101001].

\bibitem{Frezzotti:2003xj}
  R.~Frezzotti and G.C.~Rossi,
  Nucl.\ Phys.\ Proc.\ Suppl.\  {\bf 128} (2004) 193, 
  [hep-lat/0311008].

\bibitem{Frezzotti:2003ni}
  R.~Frezzotti and G.~C.~Rossi,
  JHEP {\bf 0408} (2004) 007
  [hep-lat/0306014].

\bibitem{Frezzotti:2004wz}
  R.~Frezzotti and G.~C.~Rossi,
  JHEP {\bf 0410} (2004) 070,
  [hep-lat/0407002].

\bibitem{Osterwalder:1977pc}
  K.~Osterwalder and E.~Seiler,
  Ann. Phys.\  {\bf 110} (1978) 440.

\bibitem{Jegerlehner:1996pm}
  B.~Jegerlehner,
  arXiv:hep-lat/9612014.

\bibitem{Jansen:2005kk}
  K.~Jansen, M.~Papinutto, A.~Shindler, C.~Urbach and I.~Wetzorke  [XLF Coll.],
  JHEP {\bf 0509} (2005) 071, 
  [arXiv:hep-lat/0507010].

\bibitem{McNeile:2006bz}
  C.~McNeile and C.~Michael [UKQCD Coll.],
  Phys.\ Rev.\  D {\bf 73} (2006) 074506, 
  [hep-lat/0603007].

\bibitem{Bhattacharya:2015rsa}
  T.~Bhattacharya, V.~Cirigliano, R.~Gupta, E.~Mereghetti and B.~Yoon,
  Phys.\ Rev.\ D {\bf 92} (2015) no.11,  114026
  [arXiv:1502.07325 [hep-ph]].

\bibitem{Testa:1998ez}
  M.~Testa,
  JHEP {\bf 9804} (1998) 002, 
  [hep-th/9803147].

\bibitem{Maiani:1991az}
  L.~Maiani, G.~Martinelli and C.~T.~Sachrajda,
  Nucl.\ Phys.\ B {\bf 368} (1992) 281.

\bibitem{Gabrielli:1990us}
  E.~Gabrielli, G.~Martinelli, C.~Pittori, G.~Heatlie and C.~T.~Sachrajda,
  Nucl.\ Phys.\ B {\bf 362} (1991) 475.
 
\bibitem{FLAG}
  S.~Aoki {\it et al.},
  Eur.\ Phys.\ J.\ C {\bf 77} (2017) no.2,  112
  [arXiv:1607.00299 [hep-lat]].

  See also S.~Aoki {\it et al.},
  Eur.\ Phys.\ J.\ C {\bf 74} (2014) 2890
  [arXiv:1310.8555 [hep-lat]].

\end{thebibliography}
\end{document}